\documentclass[preprint,12pt,authoryear]{elsarticle}

\usepackage{amsmath}
\usepackage{textcomp}
\usepackage{graphicx}
\usepackage{lineno}
\usepackage{soul,color}

\usepackage{setspace}

\graphicspath{{./}{figures/}}

\journal{Meteoritics and Planetary Science}

\begin{document}

\doublespacing
\sloppy

\begin{frontmatter}

\title{Oxygen Isotopic Composition of an Enstatite Ribbon of Probable Cometary Origin}


\newcommand{\washu}{Department of Physics, Washington University in St.\ Louis, St.\ Louis, MO 63130, USA}

\newcommand{\uw}{Department of Astronomy, University of Washington, Seattle, WA 98195, USA}

\newcommand{\uh}{Hawai`i Institute of Geophysics and Planetology, University of Hawai`i at M\={a}noa, Honolulu, HI 96822, USA}

\author{Ryan C. Ogliore}
\ead{rogliore@physics.wustl.edu}
\address{\washu}

\author{Donald E. Brownlee}
\address{\uw}

\author{Kazuhide Nagashima}
\address{\uh}

\author{Dave J. Joswiak}
\address{\uw}

\author{Josiah B. Lewis}
\address{\washu}

\author{Alexander N. Krot}
\address{\uh}

\author{Kainen L. Utt}
\address{\washu}

\author{Gary R.\ Huss}
\address{\uh}

\begin{abstract}
Filamentary enstatite crystals are found in interplanetary dust particles of likely cometary origin but are very rare or absent in meteorites. Crystallographic characteristics of filamentary enstatites indicate that they condensed directly from vapor. We measured the O isotopic composition of an enstatite ribbon from a giant cluster interplanetary dust particle to be $\delta^{18}\rm{O}{=25{\pm}55}$, $\delta^{17}\rm{O}{=-19{\pm}129}$, $\Delta^{17}\rm{O}{=-32{\pm}134}$ (2$\sigma$ errors), which is inconsistent at the 2$\sigma$ level with the composition of the Sun inferred from the Genesis solar wind measurements. The particle's O isotopic composition, consistent with the terrestrial composition, implies that it condensed from a gas of non-solar O isotopic composition, possibly as a result of vaporization of disk region enriched in $^{16}$O-depleted solids. The relative scarcity of filamentary enstatite in asteroids compared to comets implies either that this crystal condensed from dust vaporized \textit{in-situ} in the outer Solar System where comets formed, or it condensed in the inner Solar System and was subsequently transported outward to the comet-forming region.
\end{abstract}


\end{frontmatter}

\section{Introduction} \label{sec:intro}
The interstellar silicate absorption band around 10~$\mu$m is smooth and featureless, implying the interstellar crystalline silicate fraction is $<$2.2\% \citep{kemper2004absence}. However, even the youngest accretionary disks around new stars ($\leq 1$~Myr) show crystalline silicate fractions of 10--20\%, likely from thermal processing of dust during the first 1~Myr of disk evolution \citep{oliveira2011evolution}. 

Crystalline silicates have been observed in both the inner and outer regions of spatially resolved protoplanetary disks \citep{van2004building,riaz2012radial}. The pyroxene (nearly pure enstatite, MgSiO$_3$) to olivine (nearly pure forsterite, Mg$_2$SiO$_4$) ratio can be different in the outer and inner disk \citep{van2004building,bouwman2008formation}, which may reflect different formation conditions \citep{roskosz2011sharp} in the inner and outer disk, or annealing of amorphous precursors with different stoichiometry. \citet{murata2009crystallization} proposed that the radial dependence of the pyroxene/olivine ratio can be attributed to  differences in crystallization activation energy of amorphous silicates with different compositions --- the higher temperatures in the inner disk are required to crystallize the amorphous precursor of pyroxene. Observations of variable olivine/pyroxene ratios in the inner and outer parts of protoplanetary disks is evidence for local (in-situ) formation of these crystalline phases, rather than a scenario in which all crystalline silicates formed close to the central star and were subsequently transported to the outer disk \citep{bouwman2008formation}.

Chondritic-porous interplanetary dust particles (IDPs) and ultra-carbonaceous Antarctic micrometeorites (UCAMMs) \citep{duprat2010extreme,dartois2018dome} from the Solar System have carbon contents comparable to the C-rich dust from comet 67P/Churyumov-Gerasimenko measured by the COSIMA instrument during the Rosetta mission \citep{bardyn2017carbon}, which is significantly higher than C contents in carbonaceous chondrites \citep{thomas1993carbon}. Large ``giant cluster'' IDPs (GC-IDPs) are found to have disaggregated into hundreds to thousands of fragments upon impacting silicone oil collectors at 200~m/s during collection of stratospheric dust by high-altitude aircraft. Giant cluster IDPs are highly porous \citep{rietmeijer1993size}, fragile, mineralogically diverse, unequilibrated, and finer-grained compared to the most primitive chondrites \citep{bradley2003interplanetary,joswiak2017refractory}, implying that these particles were kept cold since their formation and were unaffected by the metamorphic and aqueous processes that lithified asteroidal rock. Comets, kept in cold storage in the outer Solar System since their formation, are known to be the source of the majority of interplanetary dust in the Solar System \citep{nesvorny2010cometary}. Other evidence that GC-IDPs are derived from comets includes \citep{joswiak2017refractory}: 1) Rocky samples of comet Wild~2 returned from the Stardust mission are more mineralogically diverse and unequilibrated than chondrites, but similar to GC-IDPs; 2) Comet Wild~2 and GC-IDP olivines show uncorrelated Mn and Fe contents, whereas asteroid samples (meteorites) tend to show strong correlations \citep{brownlee2017diversity,frank2014olivine}; 3) Kool (kosmochloric Ca-rich pyroxenes and FeO-rich olivines) grain assemblages and filamentary enstatite are present in comet Wild~2 and GC-IDP samples but are absent or exceedingly rare chondrites \citep{joswiak2009kosmochloric,stodolna2014characterization}; 4) Dust ejected from comet 67P/Churyumov-Gerasimenko and collected at low-speed by Rosetta's COSIMA instrument appeared to be fragile aggregates of fine particles, similar to the structure of GC-IDPs \citep{merouane2016dust}.

Pyroxene grains in GC-IDPs of probable cometary origin are often found in platelet, whisker, and ribbon morphologies of nearly Fe-free clinoenstatite \citep{bradley1983pyroxene}. An enstatite whisker was identified in the Stardust samples from comet Wild~2 \citep{stodolna2014characterization}. Although filamentary enstatite is commonly observed in transmission electron microscopy studies of ultramicrotomed thin sections of chondritic IDPs \citep{bradley1988analysis}, there are only a few reports of enstatite whiskers in primitive meteorites. Enstatite whiskers were reported in the CM carbonaceous chondrite Paris \citep{leroux2015gems}. However, these crystals are not perfect analogs to enstatite whiskers in IDPs since they do not show the extreme elongations seen in enstatite whiskers from GC-IDPs \citep{leroux2015gems}. An elongated enstatite grain in the matrix QUE 99177 (CR2) was reported by \citet{alexander2017nature}, though detailed crystallography is needed to compare to filamentary enstatite in IDPs. Filamentary crystals $<10$~$\mu$m long have been shown to tolerate $\sim$500 times larger strains than equiaxial crystals of the same type \citep{brenner1956growth}. Filamentary enstatite likely would have survived incorporation into asteroids, as well as the mild thermal metamorphism or aqueous alteration experienced by primitive chondrites on their parent bodies \citep{donn1963planets} (micrometer-scale crystalline fines, including pyroxenes, are found in the matrix of primitive chondrites \citep[e.g.,][]{abreu2011deciphering}). An extensive search for filamentary enstatite in disaggregated primitive chondrites (where elongated structures are more easily identified) are needed to make a more precise assessment, but from previous observations it appears that filamentary enstatite is much more abundant in rocky material from comets than in asteroids.

Filamentary crystal structures were predicted to condense directly from gas in the solar nebula and become incorporated into comets long before such crystals were observed in IDPs \citep{donn1963planets}. Nucleii with screw dislocations can grow one-dimensionally from a supersaturated vapor and outcompete equiaxial crystal growth \citep{donn1963planets}. Enstatite whiskers and ribbons in IDPs are elongated along the crystallographic [$100$] axis \citep{bradley1983pyroxene}.  Rock-forming terrestrial and meteoritic pyroxenes differ from cometary filamentary enstatite in the following ways: they are elongated along [$001$] (when not equiaxial), show twinning, or are intergrowths of ortho- and clinoenstatite. Filamentary enstatites in IDPs frequently contain axial screw dislocations and ($100$) stacking faults that are a result of their vapor-phase growth \citep{bradley1983pyroxene} and were not erased by subsequent thermal annealing. 

The O isotopic composition of cometary filamentary enstatite is a measure of the O isotopic composition of the gas from which these grains condensed. Calcium-aluminum-rich inclusions (CAIs) and amoeboid olivine aggregates (AOAs) are thought to have formed by evaporation and condensation processes in a gas of approximately solar composition in the innermost part of the Solar System at the very beginning of its evolution, 4.567 Ga \citep{connelly2012absolute}. Fine-grained, spinel-rich CAIs in primitive meteorites are inferred to be direct gas-to-solid condensates based on their volatility-fractionated (Group II) rare earth element patterns \citep{davis1979condensation}, and thus record the O composition of inner nebula gas. Primitive CAIs that escaped subsequent alteration and O-isotope exchange are enriched in $^{16}$O relative to $^{17}$O and $^{18}$O compared to the Earth ($\delta^{17}$O$ \approx -45$\textperthousand, $\delta^{18}$O$ \approx -42$\textperthousand, \citep[e.g.,][]{makide2009oxygen}), though rare exceptions exist \citep{krot2017high}. These CAIs are only slightly heavier than the $^{16}$O-rich composition of the Sun inferred from measurements of solar wind returned by the NASA Genesis mission ($\delta^{17}$O=$-
59\pm10$\textperthousand, $\delta^{18}$O$=-59\pm6$\textperthousand\  (2$\sigma$ uncertainties), \cite{mckeegan2011oxygen}). Filamentary enstatite condensing from a gas of solar composition (formed by evaporation of disk regions characterized by a dust-to-gas ratio of $\sim$0.01) would also be $^{16}$O-rich. Non-filamentary $^{16}$O-rich enstatite has been reported in AOAs \citep{krot2005origin}, the matrix of K-chondrites \citep{nagashima2015oxygen}, and comet Wild~2 \citep{defouilloy2017origin}. 

Filamentary enstatite condensing from a gas created by the vaporization of dust-enriched regions of the protoplanetary disk would have the O isotopic composition closer to that of those solids. Primordial dust (protosolar molecular cloud dust thermally unprocessed in the protoplanetary disk) has been hypothesized to be  $^{16}$O-poor, close to the O isotopic composition of the Earth ($\delta^{17}$O$ = 0$\textperthousand, $\delta^{18}$O$ = 0$\textperthousand) if the Solar System inherited primordial dust that was isotopically distinct from primordial gas \citep{krot2010oxygen}. The alternative hypothesis is that primordial dust is $^{16}$O-rich, close to the solar composition, but was intimately mixed water ice that was very $^{16}$O-poor (up to $\delta^{17}$O=$+180$\textperthousand, $\delta^{18}$O=$+180$\textperthousand, \citet{sakamoto2007remnants}), most likely due to CO self-shielding and subsequent thermal processing during T-Tauri events at the earliest stages of Solar System evolution \citep{alexander2017measuring}). In this scenario, the dust in the disk and ice together are close to the terrestrial O composition ($^{16}$O-poor). However, \textit{either} scenario predicts that solids condensing from a gas produced by evaporation of dust-rich regions (e.g., a dust-to-gas ratio of $\sim$10 (10$^3\times$ solar), similar to the chondrule-formation region \citep{ebel2000condensation}) of the protoplanetary disk would have $^{16}$O-poor compositions close to $\delta^{17,18}$O$ = 0$\textperthousand, which is distinct from a solar gas: $\delta^{17,18}$O=$-59$\textperthousand.

In this paper, we describe measurements of the O isotopic composition of a filamentary enstatite crystal from a GC-IDP of likely cometary origin to determine if it condensed from a gas of solar composition (close to $\delta^{17,18}$O=$-59$\textperthousand) or a gas produced by the evaporation of a dust-rich region of the disk (close to $\delta^{17,18}$O=$0$\textperthousand)). 


\section{Sample Description}
U2-20 GCP, a giant cluster IDP, (U2-20 GCP, Figure \ref{fig:cluster}) was collected in 1980 during U2 aircraft flights above an altitude of 20~km that lasted a total of 61~hrs (these flights were part of the University of Washington program that preceded NASA's stratospheric dust collection program). The entire particle is estimated to be $\sim$150~$\mu$m in diameter before collection. U2-20 GCP is dominated by anhydrous minerals and has an approximately bulk chondritic composition. Refractory assemblages (CAIs and AOAs) occur in U2-20 GCP at the percent level \citep{joswiak2017refractory}. It is suggested that these objects formed in the CAI-forming region and were subsequently transported to the comet-forming region \citep{joswiak2017refractory}. Presolar silicate grains and $^{15}$N-rich material \citep{joswiak2017refractory} were also identified in U2-20 GCP---indicating that this aggregate contains solids that formed in diverse environments: from the inner Solar System to another star in the Galaxy. Isotopic signatures of these diverse environments were preserved in U2-20 GCP because it was not subjected to subsequent thermal metamorphism or significant aqueous alteration---these conditions are likely met in a cometary, not asteroidal, parent body. Next, we seek additional insight into the nature of U2-20 GCP's parent body by estimating the particle's crushing strength, and compare it to known asteroidal and cometary materials.

\subsection{Crushing Strength of U2-20 GCP}

A lower bound of the crushing strength of U2-20~GCP can be estimated by calculating the pressures it experienced during atmospheric entry \citep{love1991heating}. The pressure exerted on the particle as it enters the atmosphere is given by:
\begin{equation*}
    P(t) = \frac{4 \pi r \rho_{d}}{3} \left( -g\big(x(t) \big) + \frac{3 \rho_{\mathrm{atm}}\big(x(t)\big)}{4 \rho_{d} r} v(t) ^2 \right)
\end{equation*}
where $g\big(x(t)\big)$ is the acceleration due to gravity as a function of altitude, $v(t)$ is the particle's velocity, $r$ is the particle's radius, $\rho_{d}$ is the particle's density, and $\rho_{\mathrm{atm}}\big(x(t)\big)$ is the density of the atmosphere as a function of altitude.
    
Assuming the pre-collection particle is a sphere of diameter 150~$\mu$m with a density of 0.1~g/cm$^3$ \citep{rietmeijer1993size}, we estimate that U2-20~GCP experienced pressures of $\sim$40~Pa during atmospheric entry. Since U2-20 GCP survived atmospheric entry, this is a lower-bound for its crushing strength.
    
An estimate for the upper-bound of U2-20 GCP's crushing strength can be found by considering the maximum pressures experienced upon impact with the collector flag, which fragmented the particle. Previous estimates have assumed that the particle was uniformly accelerated to 200~m/s (the aircraft's airspeed) over the diameter of the particle, resulting in pressures greater than 1~MPa. To make a more accurate physical estimate, we have used fluid dynamics simulations to model the effect of aerodynamic forces exerted on the particle prior to collision with the collector flag.

Assuming that the particle was initially at rest with respect to the surrounding air, we can compute the aerodynamic forces exerted on the particle by the turbulent flow of air around the flag. These aerodynamic forces evolve following a differential equation similar to the one used to model atmospheric entry. Namely, the particle's velocity evolves via:

\begin{equation*}
    \frac{dv}{dt} = \textrm{sgn}\left[ u(t) - v(t) \right] \frac{3 \rho_{\textrm{atm}}}{8 \rho_{d} r} \left[ v(t) - u(t) \right]^2
\end{equation*}

where $u(t)$ is the velocity of the air with respect to the collector flag, calculated using the SimFlow (SimFlow v3.1) computational fluid dynamics software package to solve the Reynold's-averaged Navier-Stokes equations for an incompressible 200~m/s flow with turbulence around a 20~cm$^2$ plate.
    
These aerodynamic effects reduce the velocity of the particle relative to the collector flag prior to impact. Once the particle reaches the collector, we return to the simple model of the collision as a uniform acceleration over the particle diameter.
    
Under this treatment, we estimate that U2-20~GCP experienced pressures of 20~kPa upon collection, indicating that the crushing strength of U2-20~GCP lies between roughly 0.04~kPa and 20~kPa.
    
    The most fragile meteorites, e.g., the ungrouped carbonaceous chondrite Tagish Lake, have crushing strengths of at least 300~kPa \citep{brown2002entry,flynn2017physical}, a factor of 15 larger than the upper limit for the strength of U2-20 GCP. Cometary meteors have strengths around 1~kPa \citep{trigo2006strength}, consistent with the bounds we calculated for U2-20 GCP and much below the strengths measured in meteorites (though a few percent of cometary meteors have high strengths, and a few percent of asteroidal meteors are fragile, \citep{flynn2017physical}). \citet{hornung2016first} calculated the strength of comet 67P/Churyumov-Gerasimenko particles collected in-situ by the ROSETTA/COSIMA instrument to be several kPa, which is within the 0.04--20~kPa range we calculated for U2-20 GCP. From this crushing strength analysis, and comparison with strengths of known asteroidal and cometary objects, we conclude that the strength of U2-20 GCP most likely supports a cometary origin for this particle.

\subsection{Enstatite Ribbon in U2-20 GCP}

\begin{figure}[htpb]
\begin{center}
\includegraphics[height=0.32\textheight]{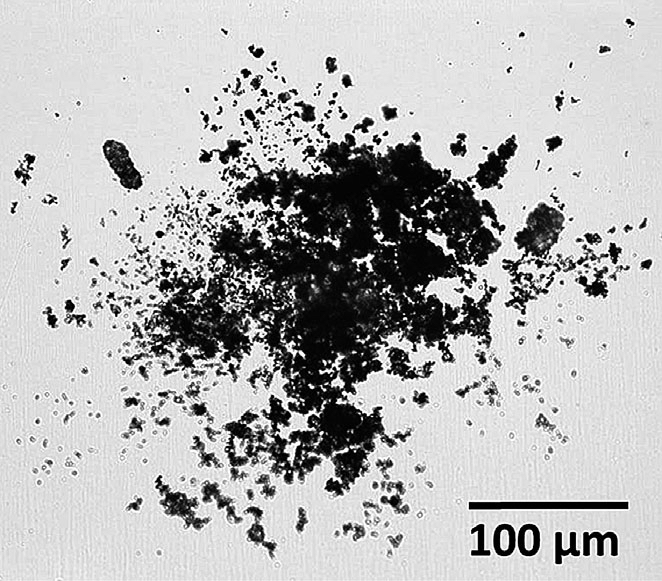}
\includegraphics[height=0.32\textheight]{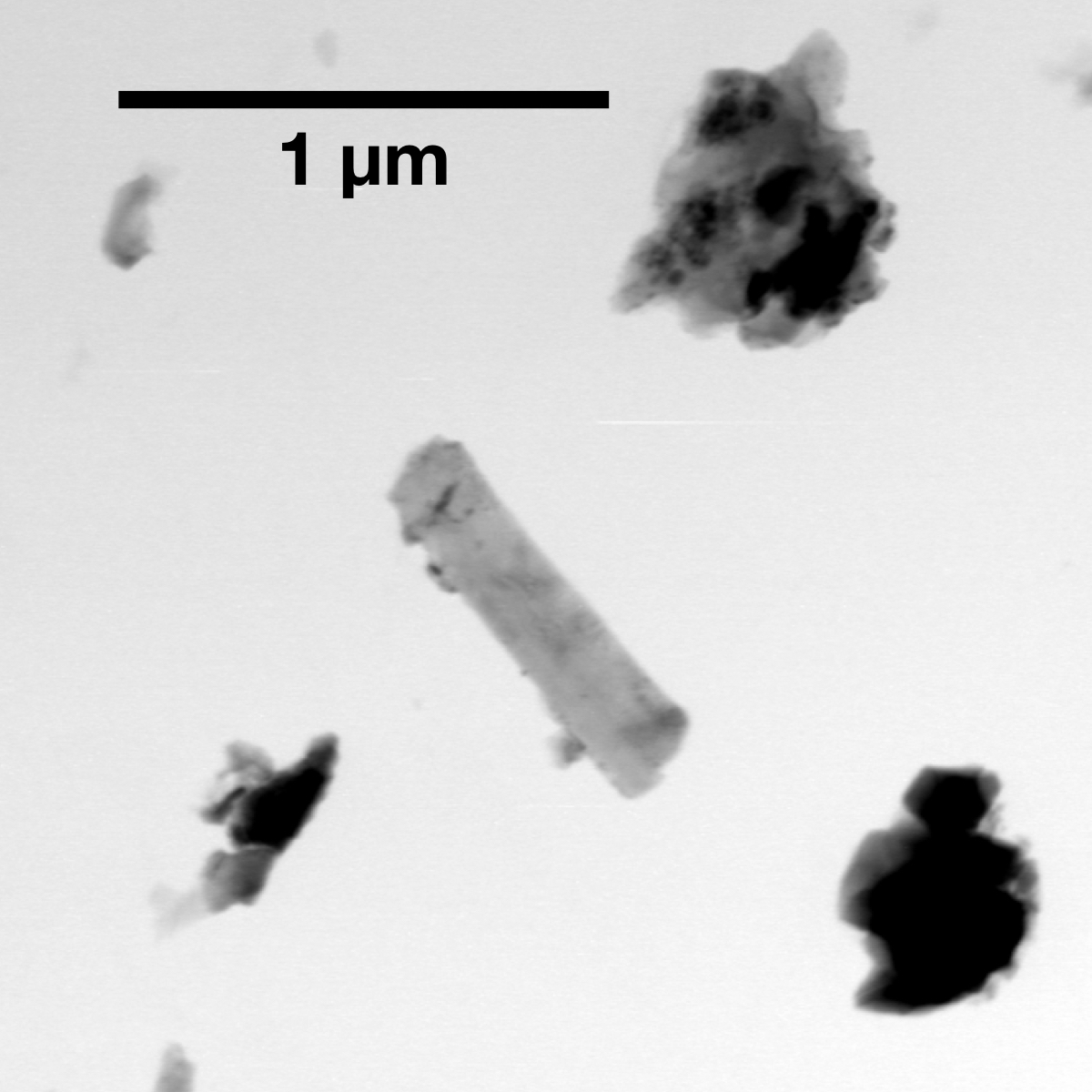}
\caption{Left) Transmitted light optical image of U2-20 GCP. Right) Bright-field TEM image of enstatite ribbon. \label{fig:cluster}} 
\end{center}
\end{figure}

We identified an 800$\times$200~nm ribbon in U2-20 GCP (Fig.\ \ref{fig:cluster}). The elemental composition of the ribbon was measured by an energy dispersive X-ray spectrometer attached to a scanning electron microscope. The ribbon was found to be consistent with Fe-poor enstatite (Mg/Si $\approx$ 1, Fe/(Mg+Fe) $<$ 0.05). This particle was not analyzed by high-resolution TEM techniques that could confirm details of its crystal structure (e.g., axial screw dislocations), but its composition and crystal habit are strong evidence it belongs to the class of filamentary enstatite with ribbon morphologies seen in interplanetary dust particles as described by \citet{bradley1983pyroxene}.

\section{Sample Preparation and Oxygen Isotope Measurements}

\begin{figure}[htpb]
\begin{center}
\includegraphics[width=\textwidth]{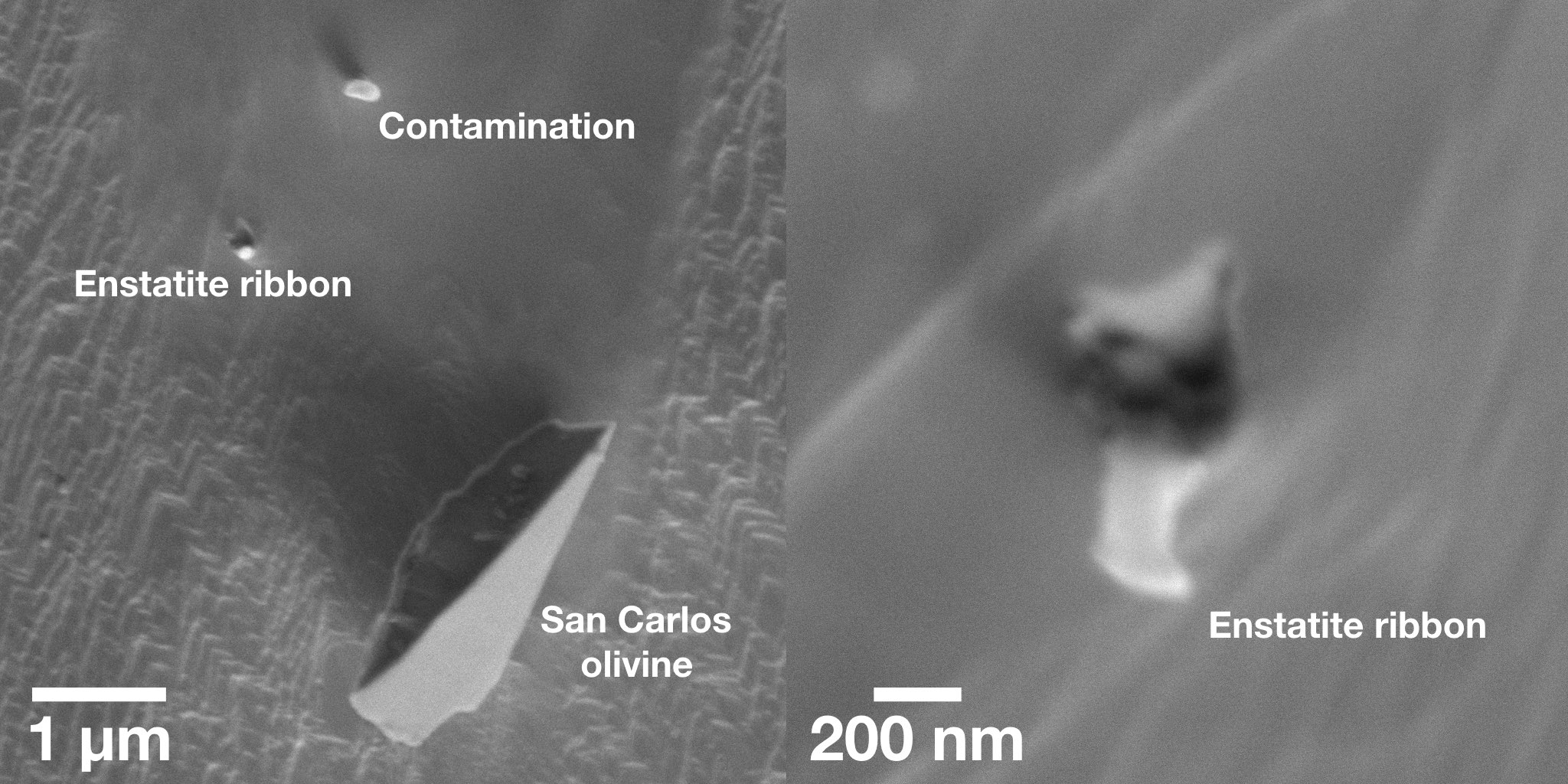}
\vspace{-2pc}
\caption{Secondary electron images of the enstatite ribbon mounted for O isotope analyses. The ribbon stuck to the Au mount nearly perpendicular to its surface.  \label{fig:mountedribbon}} 
\end{center}
\end{figure}

We transfered the enstatite ribbon from the TEM grid to a sputter-cleaned Au ion-probe mount using a computer-controlled Omniprobe needle in an FEI Quanta 3D FIB. We also transferred crushed grains of San Carlos olivine, an oxygen-isotope standard, to within 10~$\mu$m of the enstatite ribbon so that both the standard and unknown could be measured simultaneously in one raster ion image (Fig.\ \ref{fig:mountedribbon}). This method of sample preparation and analysis, simultaneous measurement of unknown and standard, will allow us to nearly eliminate instrumental mass fractionation and allow for accurate measurements of both $\delta^{18}$O and $\delta^{17}$O.

We acquired 10$\times$10~$\mu$m, 256$\times$256 pixel scanning ion images using the Cameca ims 1280 ion microprobe at the University of Hawai`i. We used a $<$3~pA Cs$^+$ primary beam focused to $\sim$250~nm. An electron flood gun was used for charge compensation. We simultaneously collected $^{16}$O$^-$, $^{17}$O$^-$, and $^{18}$O$^-$ on separate electron multipliers. Mass-resolving power for $^{17}$O$^-$ was $\sim$5500 to minimize contribution from $^{16}$OH$^-$. We also measured the $^{16}$OH$^-$ signal with electrostatic deflection (DSP2-x) to quantify any contribution of this interference to $^{17}$O$^-$ (measured to be $<1$\textperthousand). We collected 2000 scanning ion image frames (21 hours) of both the enstatite ribbon and the San Carlos olivine standard. Then we decreased the raster size to 2$\times$2~$\mu$m for 200 frames (2.7~hr) and collected images on only the sample, but with a higher sputter rate. Finally, we increased the raster back to 10$\times$10~$\mu$m for another 200 frames, again collecting images of both the San Carlos olivine standard and enstatite ribbon. This procedure allowed us to collect scanning ion images over the unknown and standard simultaneously throughout most of the measurement, bracketing the frames where only the unknown was measured. We continued the measurement until oxygen counts from the enstatite ribbon decrease to nearly zero, implying the particle was almost completely sputtered away. The useful yield of O from the enstatite was calculated from the total collected O counts (excluding the first cycles with adsorobed water) divided by an estimate of the number of O atoms in an 600$\times$200$\times$30~nm enstatite crystal. The useful yield was found to be $\sim$0.5\%, though this number is uncertain due to the imprecise estimate of the thickness of the enstatite (25~nm).

\begin{figure}[htpb]
\centering
\includegraphics[width=0.67\columnwidth]{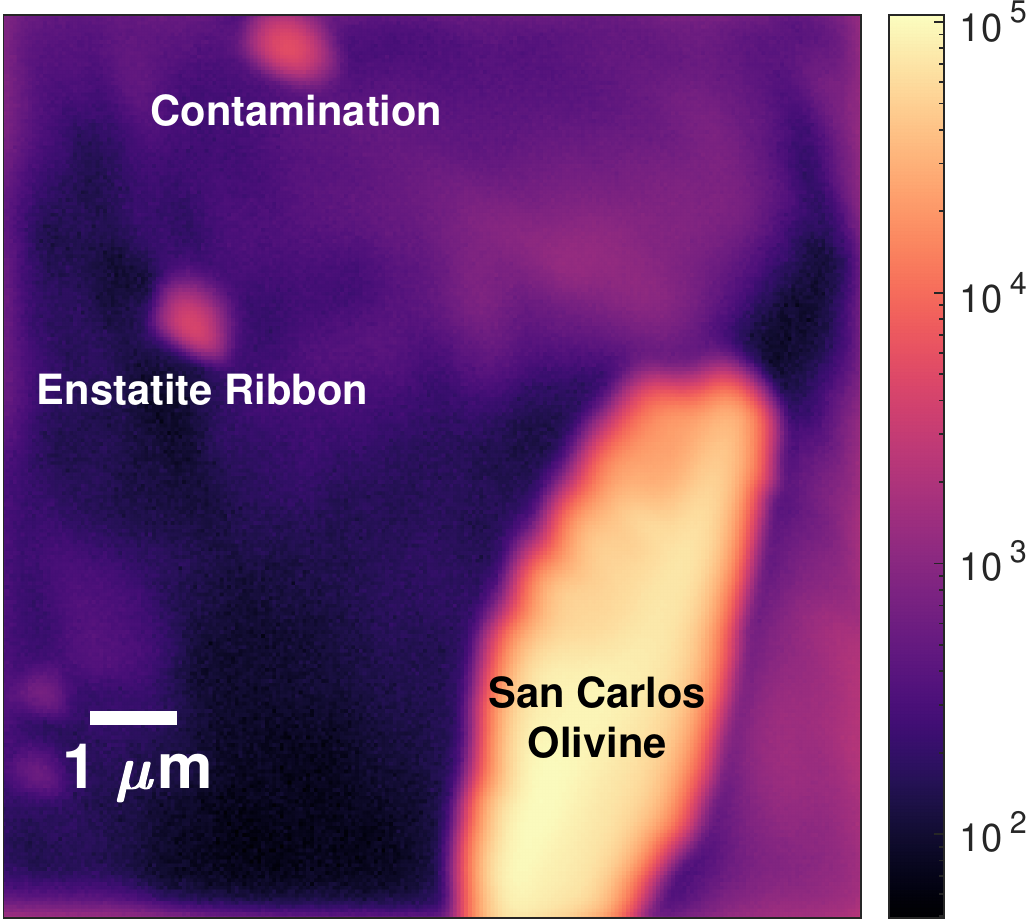}
\caption{Scanning ion image of $^{16}$O (counts per second).  \label{fig:sii}} 
\end{figure}

\section{Data Analysis}
We aligned the stack of 2400 isotope images for drift during the measurement, removed spurious measurement cycles, corrected for electron-multiplier deadtime (66 ns for multicollector, 30~ns for monocollector), then defined regions-of-interests around the enstatite whisker and San Carlos olivine. We avoided grain edges when defining the regions-of-interest to minimize instrumental mass fractionation from topography. Electron multiplier and contamination background was calculated from the data set in a region of empty gold foil away from the samples and found to be sufficiently low as to not affect the results.  We used San Carlos olivine to constrain the yields on the electron multipliers (EMs) used to measure $^{17}$O$^-$ and $^{18}$O$^-$. There was no significant change in the EM efficiency over the course of the measurement, as monitored by simultaneous measurements of the San Carlos olivine standard. We did not see any change in the efficiency in any of the EMs over the course of the measurement, likely because the count rates were relatively low ($<$10$^5$ cps). We eliminated the first 150 cycles during which adsorbed water (background O) on the Au mount was removed. The remaining 2250 cycles were used to calculate the oxygen isotopic composition of the enstatite whisker and simultaneously measured San Carlos olivine oxygen isotope standard from the regions-of-interest defined on these two objects. The enstatite whisker was well-resolved spatially from both the San Carlos olivine and contamination. 

We calculated uncertainties by a bootstrap Monte Carlo method. We randomly resampled pixels in the enstatite ribbon region-of-interest 10$^4$ times (allowing repeats) and calculated the full covariance matrix (variances and covariances) of the $\delta^{18}$O and $\delta^{17}$O values for the trials. The error ellipse calculated from the covariance matrix is shown in Fig.\ \ref{fig:oisotopes}. The uncertainties in $\delta^{18}$O and $\delta^{17}$O are calculated as the projection of this ellipse onto each axis. Monte Carlo uncertainties were $\sim$25\% larger than purely statistical uncertainties, and also larger than the reproducibility of the simultaneously measured San Carlos olivine over the 2250 cycles of the measurement which was close to statistical. For these reasons we believe our bootstrap uncertainties are the most conservative possible estimate of the measurement uncertainty. 

To quantify the instrumental mass fractionation resulting from the height difference between the San Carlos olivine surface-mounted grain and the enstatite ribbon, we measured $\sim$10~$\mu$m San Carlos olivine surface-mounted grains and polished San Carlos olivine. To obtain a precise estimate of instrumental mass fractionation between these two mounts, we used a higher primary beam current (and spot size), and collected $^{16}$O on the L1 Faraday cup (analytical conditions are similar to \citet{ogliore2012incorporation}). The mean of the measurements of the polished San Carlos olivine were constrained to the true values $\delta^{18}$O=5.25$\pm$0.9, $\delta^{17}$O=2.73$\pm$0.5, where the errors are two standard errors of the set of measurements. The surface-mounted grains were measured to have values in the range of $\delta^{18}$O=2.4--6.6 and $\delta^{17}$O=1.6--2.6. Conservatively, we estimate the maximum plausible difference between measurements of these two samples with different topography to be 4\textperthousand\ in $\delta^{18}$O and 2\textperthousand\ in $\delta^{17}$O. We added 4\textperthousand~in quadrature to our 2$\sigma$ uncertainty in $\delta^{18}$O and 2\textperthousand~in quadrature to the uncertainty in $\delta^{17}$O to account for the uncertainty due to instrumental mass fractionation caused by the height difference between the enstatite ribbon and San Carlos olivine standard.

We also considered electron-multiplier background and quasi-simultaneous arrival effects. We measured electron multiplier backgrounds to be 1$\times$10$^{-3}$ counts per second. This background correction would lower our reported $\delta^{18}$O by $\sim$0.7\textperthousand\ and $\delta^{17}$O by $\sim$2\textperthousand\, which is much smaller than our final reported uncertainty, so detector background can be ignored. Quasi-simultaneous arrival \citep{slodzian2004qsa} can cause relative undercounting of the most abundant isotope, e.g., $^{16}$O, when ion-counting detectors such as electron multipliers are used. Our secondary to primary count rate ratio was $0.003$--$0.010$ for the San Carlos olivine, with a predicted correction of $1$--$5$\textperthousand\ to the unknown based on the statistical assumptions described in \citet{slodzian2004qsa}. Since this value is also much smaller than our final reported uncertainty, the QSA effect can be ignored.

\section{Results}
\begin{figure}[htpb]
\centering
\includegraphics[width=\columnwidth]{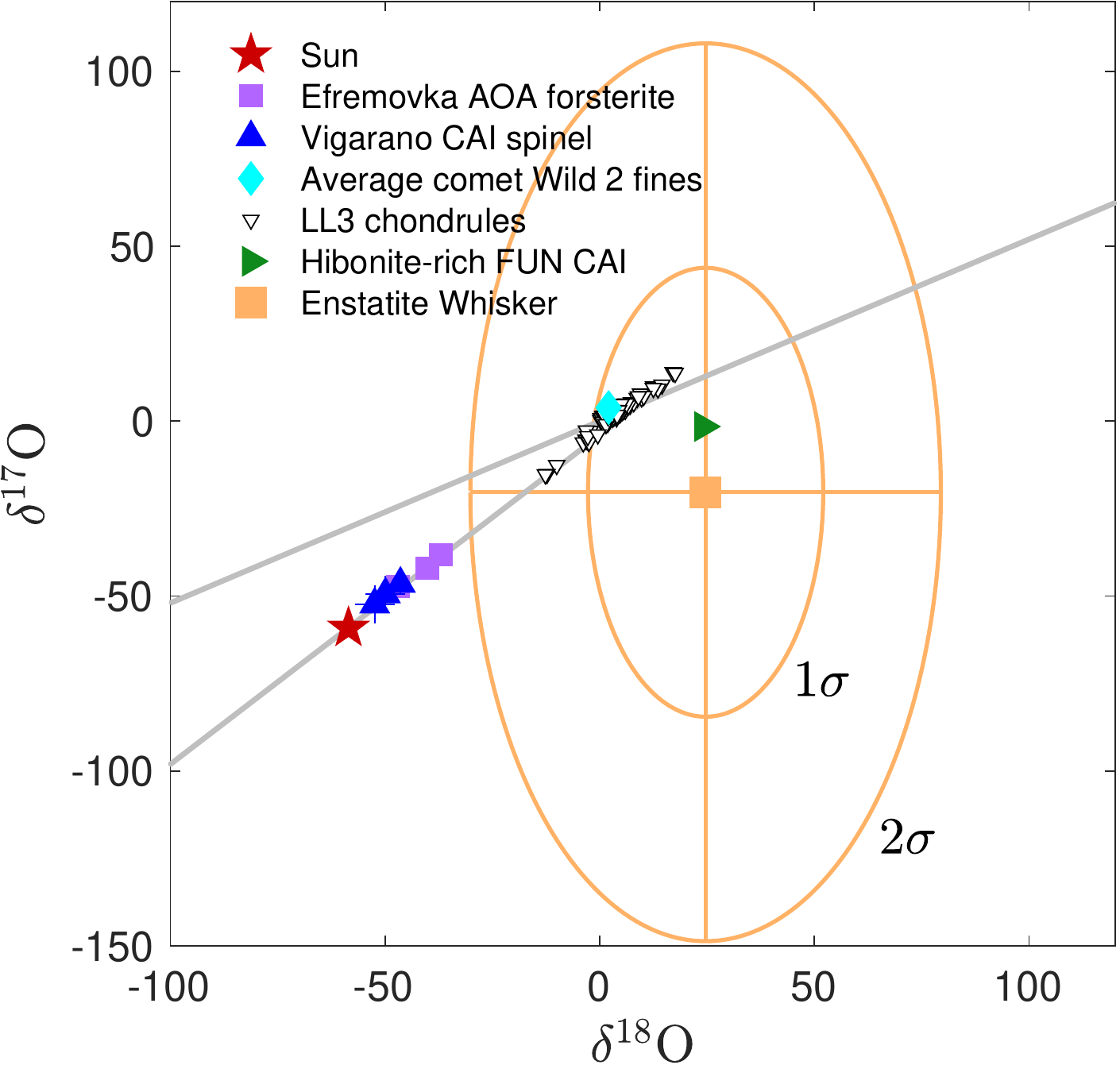}
\caption{Oxygen isotopic composition of enstatite ribbon (orange) with 1$\sigma$ and 2$\sigma$ uncertainties compared to the Sun \citep{mckeegan2011oxygen}, Efremovka AOA \citep{krot2002existence}, Vigarano CAI \citep{krot2002existence}, comet Wild~2 fines \citep{ogliore2015oxygen}, and LL3 chondrules \citep{kita2010high}. Terrestrial fractionation line and slope-one line are shown in gray. \label{fig:oisotopes}} 
\end{figure}

We measured the O isotopic composition of the enstatite ribbon to be: $\delta^{18}\rm{O}{=25{\pm}55}$, $\delta^{17}\rm{O}{=-19{\pm}129}$, $\Delta^{17}\rm{O}{=-32{\pm}134}$ (all errors 2$\sigma$). The enstatite ribbon is consistent with $^{16}$O-poor compositions seen in chondrules (formed by melting of dust in dust-rich regions of the protoplanetary disk) but is not consistent, at the 2$\sigma$ level, with the $^{16}$O-rich compositions seen in unaltered gas-to-solid condensates (CAIs and AOAs) that formed in the inner Solar System (Fig.\ \ref{fig:oisotopes}).

\section{Discussion}
Our sample preparation and analytical protocol allowed for the accurate determination of $\delta^{17}$O and $\delta^{18}$O in the cometary enstatite ribbon, which are not consistent with three-isotope O composition of the Sun \citep{mckeegan2011oxygen} and the most $^{16}$O-rich objects in chondrites at the 2$\sigma$ level. Next, we consider the possibility that the enstatite ribbon actually formed with solar $\delta^{17}$O and $\delta^{18}$O values ($\delta^{17,18}$O=$-59$\textperthousand) and was subsequently altered to its measured composition.

IDP U2-20 GCP is anhydrous so the filamentary enstatite ribbon likely did not exchange O isotopes with a water reservoir on its parent body. Additionally, U2-20 GCP is an unequilibrated collection of minerals with disparate compositions, so the filamentary enstatite likely did not exchange O isotopes with neighboring grains during thermal metamorphism on its parent body.

In situ measurements of components in meteorites plot along a slope $\approx$1 line in the three oxygen isotope plot (e.g., \citet{ushikubo2012primordial}) as shown in Figure\ \ref{fig:oisotopes}. Large mass-dependent fractionation effects (along a slope $\approx$0.5 line in Fig.\ \ref{fig:oisotopes}), up to $+40$\textperthousand\ in $\delta^{18}$O, have been measured in FUN CAIs \citep{krot2014calcium}. The mass-dependent fractionation in FUN CAIs is thought to originate from preferential loss of $^{16}$O during melt evaporation. Our measurement could be explained if the filamentary enstatite formed with solar O isotope composition followed by mass-dependent fractionation to its measured composition. However, crystallographic defects seen in other filamentary enstatites from IDPs \citep{bradley1983pyroxene} are evidence that these grains avoided thermal annealing, including the heating events experienced by FUN CAIs. Though we did not look for crystallographic defects in the En ribbon reported on here, its crystal habit, composition, and nature of its host particle indicate that this En ribbon most likely experienced a similar history as other filamentary enstatites found in cometary IDPs. Mass-dependent fractionation from post-crystallization heating, therefore, is unlikely to explain its O isotope composition. Oxygen isotope exchange between a precursor $^{16}$O-rich melt and a $^{16}$O-poor gas \citep{kita2010high} may be able to explain our measured O composition. However, since the En ribbon is likely a gas-to-solid condensate, and never existed as a melt, we rule out this scenario. Finally, we consider the scenario where the En ribbon condensed from a $^{16}$O-rich gas and later experienced O isotope exchange with $^{16}$O-poor gas phases (e.g., H$_2$O) in the protoplanetary disk \citep{yamamoto2018oxygen}. While amorphous silicates can efficiently exchange O isotopes with the gas above 600~K, if crystallization precedes O isotope exchange, this process becomes very inefficient \citep{yamamoto2018oxygen}. For filamentary entastite grains that likely condensed as crystalline phases directly from the gas, their original O isotopic compositions are likely preserved since there was no significant O isotope exchange with gas in the protoplanetary disk.


Our measurements indicate that this cometary enstatite ribbon most likely condensed from a $^{16}$O-poor gas. CO self-shielding can create $^{16}$O-poor gas from a primordial $^{16}$O-rich reservoir, but this gas-phase O is converted to H$_2$O quickly through interaction with nebular H ($\sim$10$^5$ years, \cite{lyons2005co}). The $^{16}$O-poor gas composition sampled by the cometary enstatite ribbon we measured was more likely created by vaporization of $^{16}$O-depleted solids. The formation of this enstatite ribbon by vaporization-recondensation could have happened 1) \textit{in-situ} in the outer Solar System where its cometary parent body formed or 2) in the inner Solar System followed by transport to the outer Solar System where it was incorporated into its parent comet. We consider these two scenarios in detail below.

\paragraph{Outer Solar System formation:} 


Gravitational instabilities in protoplanetary disks create global spiral density waves which grow and produce shocks in the disk \citep{cassen1981numerical}. Models predict that shock processing of dust beyond $\sim$10~AU becomes difficult due to lower gas densities \citep[e.g.,][]{harker2002annealing}. Recently, however, spiral density waves around the protoplanetary disk of the young star Elias 2-27 were observed to extend to the outer disk \citep{perez2016spiral}. The combination of gravitational instabilities and planet-disk interactions may cause spiral arms to extend to the outer regions of protoplanetary disks \citep{pohl2015scattered}. A very early formation of Jupiter by rapid accretion of its 20M$_\oplus$ rocky core \citep{pollack1996formation,warren2011noncarb,kruijer2017age} may have contributed to the presence of extended spiral arms and dust evaporation by shock heating in the outer Solar System. 

Another mechanism to vaporize dust in the outer Solar System is via the formation and destruction of planetary embryos in the young solar nebula, as described by  \citet{vorobyov2011destruction}. This mechanism was suggested by \citet{bridges2012chondrule} to explain the abundance of high-temperature objects such as chondrule fragments in the Stardust samples returned from comet Wild~2. The contraction of planetary embryos creates gas temperatures in their interiors high enough to vaporize dust \citep{vorobyov2011destruction}. Disruption of these embryos by tidal forces, as they migrate inward from $>$100~AU \citep{vorobyov2006burst}, can deposit their constituent gas and dust (including crystalline silicates, especially robust crystals like filamentary enstatite) in the outer Solar System.

\paragraph{Inner Solar System formation}
The vaporization of dust is more plausible in the inner solar nebula where gas densities and pressures are higher. For example, primordial silicate dust grains can be heated to vaporization as they pass through the shock created by infalling gas at the surface of the proto-Sun \citep[e.g.,][]{ruzmaikina1994chondrule,chick1997thermal}, which may also explain the crystalline fractions of dust in the inner regions of protoplanetary disks \citep{van2004building}.  

If filamentary enstatite formed in the inner Solar System, one would expect these crystals to be at least as abundant in meteorites as they are in cometary materials. The abundance of refractory inclusions, including CAIs which formed close to the Sun, in cometary material (sampled from two comets) is about equal to that in carbonaceous chondrites \citep{joswiak2017refractory}. Filamentary enstatite appears to be far more abundant in cometary material than in asteroidal material, though a thorough statistical study has not been done. This observation can be explained if filamentary particles, with very large surface area to volume ratios, can more efficiently couple to the gas and be evacuated from the inner disk to the formation region of Jupiter-family comets in the outer disk, where gas density is lower. Evidence for aerodynamic grain sorting in cometary dust was reported by \citet{wozniakiewicz2012grain} in a statistical study of sizes of sulfide and silicate grains in chondritic-porous interplanetary dust particles and Stardust samples from comet Wild~2. If aerodynamic processes played a role in transporting small-grains to the comet-forming regions in the outer Solar System, filamentary enstatite grains may be preferentially transported and efficiently removed from the inner Solar System where asteroids formed.


\section*{Conclusions}

We measured an enstatite ribbon (commonly found in cometary material but exceedingly rare in meteorites) to have an O isotopic composition inconsistent (at the 2$\sigma$ level) with a formation by condensation of a gas of solar composition. Our measurement can be explained by either the condensation of this grain from $^{16}$O-depleted vaporized solids in the outer Solar System, or an inner Solar System formation followed by transport of this grain to the outer Solar System where comets formed. Additional oxygen-isotope measurements of filamentary enstatite from different giant-cluster IDPs is needed to constrain the variability of formation environments of filamentary enstatite.

An exhaustive search for filamentary enstatite (with defects that are diagnostic of vapor condensation) in disaggregated primitive chondrites will help determine if these crystals can form and be retained at all in the inner Solar System. If filamentary enstatite is identified in primitive chondrites, the O isotopic composition of these crystals will help determine if \textit{asteroidal} filamentary enstatite is related to \textit{cometary} filamentary enstatite. If asteroidal and cometary filamentary enstatite have similar O isotopic compositions (i.e.~both inconsistent with condensation from a gas of solar composition) it is likely that these asteroidal and cometary crystals are genetically related. The most likely scenario is that they both condensed in the inner Solar System but were preferentially transported to the outer Solar System to be incorporated into comets.

However, if no filamentary enstatite is found in primitive chondrites, or asteroid and cometary filamentary enstatite have different O isotopic compositions (i.e.~asteroid filamentary enstatite is consistent with condensation from a gas of solar composition), filamentary enstatite in comets is not genetically related to inner Solar System material. In this case, it is most likely that filamentary enstatite formed \textit{in situ} in the outer Solar System.

\section{Acknowledgements}
This work was supported by NASA Grant NNX14AF24G (R.~C. Ogliore, P.I.). The authors thank associate editor Larry Nittler, reviewers Shogo Tachibana and C\'{e}cile Engrand, and an anonymous reviewer for helpful suggestions that improved this manuscript. This work is dedicated to Christine Floss in appreciation of her scientific accomplishments and contributions to the fourth-floor group at Washington University in St.\ Louis.






\section{References}
\bibliographystyle{elsarticle-harv}
\bibliography{apjl}



\end{document}